\documentclass[]{raa}            
\usepackage{graphicx,times}
\usepackage{natbib}
\usepackage{amssymb}

\providecommand{\bm}[1]{\mbox{\boldmath$#1$\unboldmath}}

\def\be{\begin{equation}}
\def\ee{\end{equation}}

\catcode`\@=11 
\def\@versim#1#2{\vcenter{\offinterlineskip
        \ialign{$\m@th#1\hfil##\hfil$\crcr#2\crcr\sim\crcr } }}

\begin{document}

   \title{The effects of viscosity on the circumplanetary disks
}

 \volnopage{ }
   \setcounter{page}{1}

   \author{De-Fu Bu
      \inst{1}
   \and Hsien Shang
         \inst{2}
   \and Feng Yuan
      \inst{1}
   }

   \institute{Key Laboratory for Research in Galaxies and Cosmology,
Shanghai Astronomical Observatories, Chinese Academy of Sciences, 80
Nandan Road, Shanghai 200030, China. {\it dfbu@shao.ac.cn; fyuan@shao.ac.cn}\\
        \and
Academia Sinica, Institute of Astronomy and Astrophysics, Taipei,
Taiwan.
 \vs \no
   {\small Received [] [] []; accepted [] [] [] }
}

\abstract{The effects of viscosity on the circumplanetary disks
residing in the vicinity of protoplanets are investigated through
two-dimensional hydrodynamical simulations with the shearing sheet
model. We find that viscosity can affect properties of the
circumplanetary disk considerably when the mass of the protoplanet
is $M_p \lesssim 33M_\oplus$, where $M_\oplus$ is the Earth mass.
However, effects of viscosity on the circumplanetary disk are
negligibly small when the mass of the protoplanet $M_p \gtrsim
33M_\oplus$. We find that when $M_p \lesssim 33M_\oplus$, viscosity
can disrupt the spiral structure of the gas around the planet
considerably and make the gas smoothly distributed, which makes the
torques exerted on the protoplanet weaker. Thus, viscosity can make
the migration speed of a protoplanet lower. After including
viscosity, size of the circumplanetary disk can be decreased by a
factor of $\gtrsim 20\%$. Viscosity helps to transport gas into the
circumplanetary disk from the differentially rotating circumstellar
disk. The mass of the circumplanetary disk can be increased by a
factor of $50\%$ after viscosity is taken into account when $M_p
\lesssim 33M_\oplus$. Effects of viscosity on the formation of
planets and satellites are briefly discussed.  }

   \authorrunning{Bu et al.}            
   \titlerunning{Viscous circumplanetary disks}  
   \maketitle


%

\section{Introduction}

Up to date, more than 500 exoplanets have been detected. Most of the
exoplanets are gas giant planets, as massive planets are
preferentially observed by current detection methods. Thus, it is
important to understand the formation process of gas giant planets.
According to the core accretion model, a solid core with several
$M_\oplus$, forms first through coagulation of planetesimals in the
circumstellar disk. The protoplanet captures a hydrostatic envelope
when its mass is less than $M_p\lesssim 10M_\oplus$ (Mizuno 1980;
Stevenson 1982; Bodenheimer \& Pollack 1986; Pollack et al. 1996;
Ikoma, Nakazawa \& Emori 2000; Ikoma, Emori \& Nakazawa 2001;
Hubickyj, Bodenheimer \& Lissauer 2005). Ikoma et al. (2000) showed
that run-away gas accretion is triggered when the solid core mass
exceeds $\simeq 5-20M_\oplus$, the protoplanet quickly increases its
mass by gas accretion. A gas giant planet acquires almost all of its
mass in the run-away gas accretion phase.

Since gas accreting from a differentially rotating circumstellar
disk has nonzero angular momentum, a circumplanetary disk can form
around the protoplanet (Tanigawa \& Watanabe 2002). The
circumplanetary disk can influence many aspects of the protoplanet.
For example, previously, when calculating the torque on the
protoplanet, the contribution of gas inside the whole Roche lobe (or
Hill radius) is neglected. This may be not appropriate. The size of
the circumplanetary disk may be smaller than the Roche lobe (as
shown in Section 3.2); therefore, when calculating the torque on the
protoplanet the gas inside the Roche lobe but beyond the outer
boundary of the circumplanetary disk should be taken into account.
Also, properties of the circumplanetary disk can determine the
evolution of protoplanet and its resulting mass. Finally, satellites
form in the vicinity of the protoplanet; studying the properties of
the circumplanetary disk helps to investigate the formation process
of the satellites.

Gas accretion process onto a protoplanet have been investigated in
global simulations by many authors (e.g. Bryden et al. 1999; Kley
1999; Lubow, Seibert \& Artymowicz 1999; Kley, D'Angelo \& Henning
2001; D'Angelo, Henning \& Kley 2002; Bate et al. 2003; D'Angelo,
Kley \& Henning 2003). However, since the main purpose of these
studies was gap formation and planet migration on a large scale, the
region in the vicinity of the protoplanet has not been investigated
with sufficient resolution. Thus, in their simulations, properties
of the circumplanetary disk were not thoroughly explored. The fine
structure of the circumplanetary disk has been investigated with
shearing sheet models without viscosity (Tanigawa \& Watanabe 2002;
Machida et al. 2008; Machida 2009; Machida et al. 2010). A question
is the mechanism of angular momentum transport in the
circumplanetary disk in their models. Gravitational interaction
between the protoplanet and gas can produce spiral shocks inside the
Roche lobe (or Hill radius) of a protoplanet. Gas flows into the
Hill sphere of a protoplanet through the inner and outer Lagrange
points. The gas that flows into the Roche lobe from the inner
(outer) Lagrange point will undergo a strong shock on the opposite
outer (inner) Lagrange side of the Roche lobe, angular momentum is
lost through the collision between gas and shocks. The gas spirals
inward toward the protoplanet as a result of successive shocks.

Obviously, there should be viscosity in the circumstellar disk,
which drives the gas flow in the disk toward the central star. The
most promising origin of viscosity in the circumstellar disk should
be magnetic turbulence generated by the magnetorotational
instability (MRI) (Balbus \& Hawley 1991; 1998). Previous work found
that despite the ionization rate of a circumstellar disk is low,
magnetic field can remain dynamically important, and MRI
perturbations can grow under a wide range of fluid conditions and
magnetic field strengths (Salmeron \& Wardle 2005). Viscosity may
play important roles on the properties of the circumplanetary disk.
The angular momentum profiles of a circumplanetary disk may be
affected significantly by viscosity. Also, viscosity helps to
transport gas into the circumplanetary disk from the differentially
rotating circumstellar disk, the mass of the circumplanetary disk
may be affected by viscosity significantly. Therefore, it is of
great importance to study the effects of viscosity on properties of
circumplanetary disk.

In this paper, we study the effects of viscosity on circumplanetary
disks with the shearing sheet models. We use an anomalous stress
tensor to mimic the shear stress originated from magnetohydrodynamic
(MHD) turbulence. In Section 2, we describe our models. Results are
described in Section 3. We discuss and summarize our results in
Section 4.

\section{Model}
\subsection{Equations}
We confine our models to two-dimensions. We assume that the
temperature is constant and that the self-gravity of the disk is
negligible. The orbit of the protoplanet is assumed to be circular
in the equatorial plane of the circumstellar disk. The protoplanet
is not allowed to migrate.

We consider a local region around a protoplanet, using the shearing
sheet model (e.g. Goldreich \& Lynden-Bell 1965). We take local
Cartesian coordinates rotating with the protoplanet with the origin
at the protoplanet and the $x-$ and $y-$ axes in the radial and
azimuthal direction of the disk, respectively. We solve the
equations of hydrodynamics without self-gravity:
\begin{equation}
\frac{\partial \Sigma}{\partial t}+\nabla \cdot (\Sigma
\textbf{v})=0
\end{equation}
\begin{equation}
\frac{\partial\textbf{v}}{\partial t}+(\textbf{v} \cdot
\nabla)\textbf{v}=-\frac{1}{\Sigma}\nabla P -\nabla \Phi-2\Omega_p
\textbf{e}_z \times \textbf{v} + \frac{1}{\Sigma} \nabla \cdot {T}
\end{equation}
where $\Sigma$ is the surface density, $\textbf{v}$ is the velocity,
$P$ is the vertically integrated gas pressure, $\Phi$ is the
gravitational potential, $\Omega_p$ is the Keplerian angular
velocity of the protoplanet, $\textbf{e}_z$ is the unit vector along
the rotation axis of the protoplanet, $T$ is the vertically
integrated anomalous stress tensor. We adopt an isothermal equation
of state, \begin{equation} P=\Sigma c_s^2 \end{equation} where $c_s$
is the sound speed. The angular velocity of the protoplanet is given
by
\begin{equation} \Omega_p=\left(\frac{GM_c}{a_p^3}\right)^{1/2}
\end{equation} where $G$, $M_c$ and $a_p$ are gravitational
constant, the mass of the central star and the orbital radius of the
protoplanet, respectively. The gravitational potential is given by
\begin{equation}
\Phi=-\frac{3}{2} \Omega_p^2 x^2-\frac{GM_p}{r}
\end{equation}
where $M_p$ and $r$ are the mass of the protoplanet and the distance
from the center of the protoplanet, respectively. The first term is
composed of the gravitational potential of the central star and the
centrifugal potential. The second term is the gravitational
potential of the protoplanet. The Hill radius inside which the
protoplanet gravity dominates is defined as
\begin{equation}
R_H=\left(\frac{M_p}{3M_c}\right)^{1/3}a_p
\end{equation}
Using the Hill radius, we can rewrite equation (4) as
\begin{equation}
\Phi=\Omega_p^2 \left(-\frac{3x^2}{2}-\frac{3R_H^3}{r}\right)
\end{equation}
We use the stress tensor $T$ to mimic the shear stress, which is in
reality magnetic stress associated with magnetohydrodynamics (MHD)
turbulence driven by the magnetorotational instability (MRI). We
assume that the only non-zero component of the stress tensor $T$ is
\begin{equation}
T_{xy}=\mu \left(\frac{\partial v_y}{\partial x}+\frac{\partial
v_x}{\partial y}\right)
\end{equation} This is because MRI is driven only by the shear
associated with orbital dynamics. In equation (7), $\mu= \Sigma
\alpha c_s^2/\Omega_k$. $\alpha$, $c_s$ and $\Omega_k$ are viscosity
coefficient, sound speed and Keplerian angular velocity about the
protoplanet, respectively. Previous paper (e.g. Papaloizou, Nelson
\& Snellgrove 2004) studying the interaction between a protoplanet
and magnetized circumstellar disk found that $\alpha\sim 3\times
10^{-3}$. The maximum $\alpha$ used in this paper is $3\times
10^{-3}$.

In this paper, we normalize length by the scale height of the
circumstellar disk $h=c_s/\Omega_p$, time by the inverse of the
Kplerian angular velocity of the protoplanet $\Omega_p^{-1}$, and
the surface density by the unperturbed surface density of  the
standard solar nebular model (Hayashi 1981; Hayashi et al. 1985).

\subsection{Numerical method}
The numerical simulations are performed using the ZEUS-2D code
(Stone \& Normal 1992a; 1992b). Our initial settings are similar to
those of Machida et al. (2008). The gas flow has a constant shear in
the $x-$direction as
\begin{equation}
\textbf{v}_0=\left(0, -\frac{3}{2}x\right)
\end{equation}
Initially, the gas has uniform surface density of the unperturbed
disk. In this paper, our standard computational domain is that
$|x|\leq 6 (\equiv x_{max})$ and $|y|\leq 12 (\equiv y_{max}) $. We
adopt logarithmic spacing grids with the finest resolution ($\Delta
x=0.003$) around the protoplanet. The total number of grids for our
standard computational domain is $500\times 1000$, the resolution is
high enough to study the circumplanetary disk in the vicinity of the
protoplanet. We inject gas with the linearized Keplerian shear on
$y=y_{max}$, $(0<x<x_{max})$ and $y=-y_{max}$, $(-x_{max}<x<0)$. For
the rest of the boundaries, we adopt outflow boundary conditions.

In order to avoid singularity in the proximity of the protoplanet,
the gravitational potential of the protoplanet is smoothed in its
neighborhood with
\begin{equation}
\Phi_p=-\frac{GM_p}{(r^2+r_{sm}^2)^{1/2}}
\end{equation}
where $r_{sm}$ is the smoothing length of the protoplanet's
potential. In this paper, we choose $r_{sm}=0.05$. Tanigawa \&
Watanabe (2002) showed that $r_{sm}=0.05$ is safe to study the
properties of the circumplanetary disk in the vicinity of the
protoplanet.

In the run-away gas accretion phase, accretion by planet is
non-negligible. There should be accretion, neglecting accretion is
unphysical. Therefore, we should mimic the accretion process by
planet. Because the growth timescale of a planet is much longer than
the typical time of our simulations, we should have a constant
accretion rate which is independent from the parameters used to
mimic the accretion process. As done by Tanigawa \& Watanabe (2002),
the gas inside $r_{sink}$ is removed by a constant rate
$[\Sigma^{n+1}=\Sigma^n(1-\Delta t)]$, $\Delta t (\ll 1)$ is a time
step of the calculations, and superscript $n$ is the number of
numerical time step. Tanigawa \& Watanabe (2002) have tested the
effects of $r_{sm}$ and $r_{sink}$ on the accretion rate of the
planet. They found that the results do not depend on the values of
$r_{sm}$ and $r_{sink}$ as long as $r_{sm}=r_{sink} < 0.07$. In this
paper, we set $r_{sm}=r_{sink} = 0.05$.

\begin{figure}[h!!!]
   \centering
   \includegraphics[width=9.0cm, angle=0]{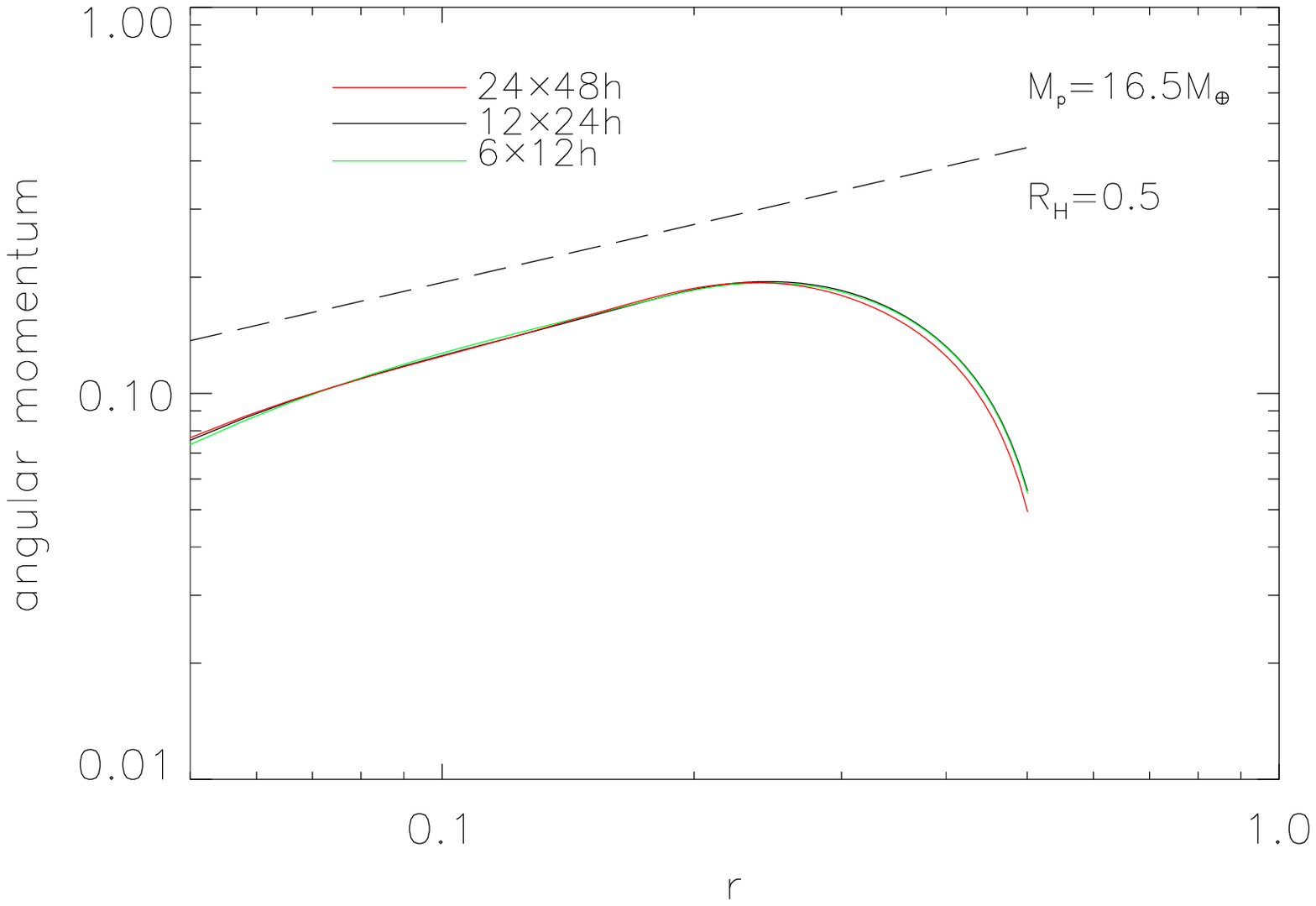}
   \includegraphics[width=9.0cm, angle=0]{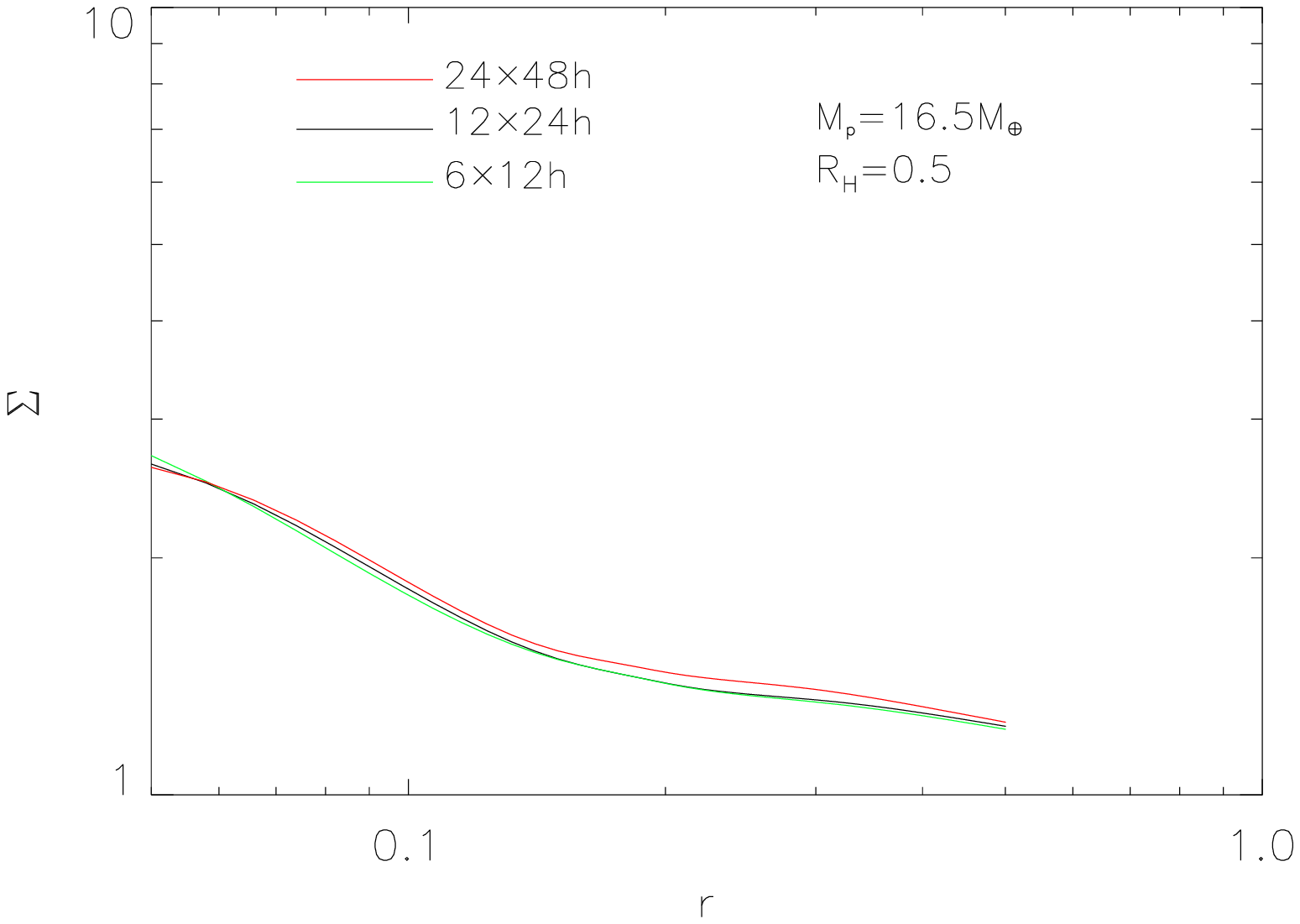}

   \begin{minipage}[]{85mm}

   \caption{Specific angular momentum (upper-panel) and surface density
(lower-panel) of the circumplanetary disks for a $16.5M_\oplus$
protoplanet as a function of distance $r$ from the protoplanet with
different sheet size. It can be seen that the effects of sheet size
on the structure of the circumplanetary disk are negligible as long
as the sheet size is much larger than the Hill radius.}
\end{minipage}
   \label{fig.gammamin}
   \end{figure}

   \begin{figure}[h!!!]
   \centering
   \includegraphics[width=9.0cm, angle=0]{dcontour-0.4.ps}
   \includegraphics[width=9.0cm, angle=0]{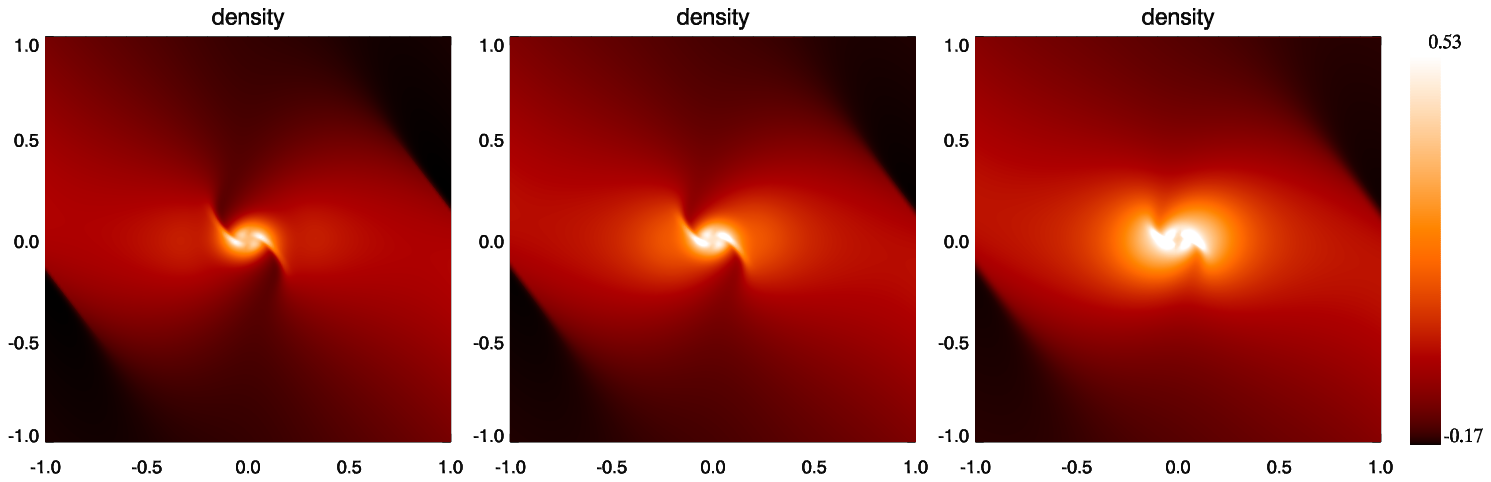}
   \includegraphics[width=9.0cm, angle=0]{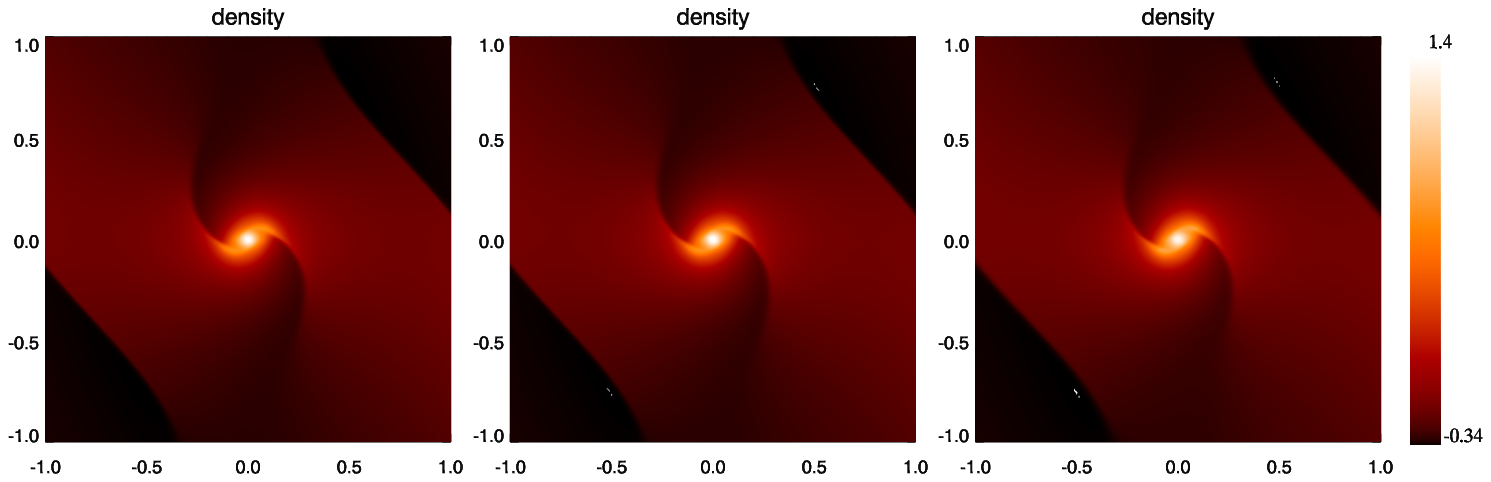}

   \begin{minipage}[]{85mm}

   \caption{Contours of logarithm surface density of the
circumplanetary disks surrounding protoplanets with mass of $8.5$,
$16.5$ and $33 M_\oplus$ (top to bottom rows). The top row panels
show the circumplanetary disks surrounding a $8.5 M_\oplus$ planet.
In the top row, the left-, middle- and right-hand panels correspond
to models M0026V0 ($\alpha=0$), M0026V1 ($\alpha=10^{-3}$) and
M0026V2 ($\alpha=3\times 10^{-3}$), respectively. The middle row
panels show the circumplanetary disks surrounding a $16.5 M_\oplus$
planet. In the middle row, the left-, middle- and right-hand panels
correspond to models M005V0 ($\alpha=0$), M005V1 ($\alpha=10^{-3}$)
and M005V2 ($\alpha=3\times 10^{-3}$), respectively. The bottom row
panels show the circumplanetary disks surrounding a $33 M_\oplus$
planet. In the bottom row, the left-, middle- and right-hand panels
correspond to models M01V0 ($\alpha=0$), M01V1 ($\alpha=10^{-3}$)
and M01V2 ($\alpha=3\times 10^{-3}$), respectively.}
\end{minipage}
   \label{fig.gammamin}
   \end{figure}

\subsection{scaling}
In the standard solar nebular model (Hayashi 1981; Hayashi et al.
1985), the temperature $T$, sound speed $c_s$, and gas density
$\rho_0$ are given by \begin {equation} T=280 \left(
\frac{L}{L_\odot} \right)^{1/4} \left( \frac{a_p}{1
AU}\right)^{-1/2}
\end {equation} where $L$ and $L_\odot$ are the protostellar and
solar luminosities;
\begin{equation}
c_s=\left( \frac{kT}{\mu m_H} \right)^{1/2}=1.9 \times 10^4 \left(
\frac{T}{10 K} \right)^{1/2} \left( \frac{2.34}{\mu} \right)^{1/2}
cm/s
\end{equation}
where $\mu=2.34$ is the mean molecular weight of the gas composed
mainly of $H_2$ and $He$; and
\begin{equation}
\rho_0=1.4 \times 10^{-9} \left( \frac{a_p}{1AU} \right)^{-11/4}
g/cm^3,
\end{equation} respectively. When the values of $M_c=1 M_\odot$ and
$L=1L_\odot$ are adopted, using equations (4), (11) and (12), we can
describe the scale height $h$ as
\begin{equation}
h=5.0 \times 10^{11} \left( \frac{a_p}{1AU} \right)^{5/4} cm.
\end{equation}

The mass of the planet in unit of Jupiter mass can be described as
\begin{equation}
\frac{M_p}{M_{Jup}}=0.12 \left( \frac{M_c}{1M_\odot} \right)^{-1/2}
\left( \frac{a_p}{1AU} \right)^{3/4} \left( \frac{R_H}{h} \right)^3
\end{equation}

In this paper, we assume that $M_c=1M_\odot$, $L=1L_\odot$. The
planet is located at $a_p=5.2 AU$. Therefore, the temperature of the
gas is $T = 123K$. Because our shearing box just represent a local
region around the planet, we further assume that the temperature is
uniform in the whole computational domain.

In the paper below, the Hill radius $R_H$ is in unit of $h$. In
table 2, $R_H=0.4$, $0.5$ and $0.63$ correspond to planet mass of
$0.026$, $0.05$ and $0.1$ Jupiter mass, respectively.

\subsection{Tests of the effects of sheet size on the results}
\begin{table}
\footnotesize
\begin{center}
\caption{Parameters for the test models}
\begin{tabular}{ccccc}
\hline \hline
Models &  Sheet size & Grids  & $R_H$ & $\alpha$\\
\hline
M005V0 & $ 12 \times 24h $ & $500 \times 1000$ & 0.5 & 0\\
BM005V0 & $ 24 \times 48h $ & $656 \times 1312$ & 0.5 & 0\\
SM005V0 & $ 6 \times 12h $ & $ 358 \times 716$  & 0.5 & 0\\
\hline
\end{tabular}
\end{center}
\end{table}

In this section, we study the effects of sheet size on the structure
of the circumplanetary disk inside the Hill radius. We find that the
flow structure inside the Hill sphere (or circumplanetary disk) does
not depend on the computational box size as long as the size is much
larger than the Hill radius. Here we just take $R_H=0.5, \alpha=0$
as a example. Table 1 lists the parameters of the $R_H=0.5$ tests.

Fig.1 shows the specific angular momentum (upper-panel) and surface
density (lower-panel) distribution of the circumplanetary disks for
a $16.5M_\oplus$ protoplanet with different sheet size. From this
figure, we can see that the effects of sheet size on the structure
of the circumplanetary disk are negligible as long as the sheet size
is much larger than the Hill radius. The results shown in Section 3
is calculated with our standard sheet size $12\times24h$.

\section{results}
Table 2 lists all of the models in this paper. We find that all of
the models have settled into their steady state before 20 orbits, so
that the results shown below at $t=75$ orbits are fully relaxed.

In order to investigate the effects of viscosity on the
circumplanetary disk, for each protoplanet with different mass, we
carry out three simulations with different viscosity coefficient
$\alpha$. We find that when the protoplanet mass $M_p \gtrsim
33M_\oplus $, the effects of viscosity are negligible.

\begin{table}
\footnotesize
\begin{center}
\caption{Parameters for all of our models}
\begin{tabular}{cccc}
\hline \hline
Models & $ {R}_H $ & $M_p$ & $\alpha$  \\
\hline
M0026V0 & $ 0.4 $ & $8.5M_\oplus$  & $0$  \\
M0026V1 & $ 0.4 $ & $8.5M_\oplus$  & $10^{-3}$  \\
M0026V2 & $ 0.4 $ & $8.5M_\oplus$  & $3\times 10^{-3}$  \\
M005V0 & $ 0.5 $ & $16.5M_\oplus$  & $0$  \\
M005V1 & $ 0.5 $ & $16.5M_\oplus$  & $ 10^{-3}$  \\
M005V2 & $ 0.5 $ & $16.5M_\oplus$  & $ 3\times 10^{-3}$  \\
M01V0 & $ 0.63 $ & $33M_\oplus$  & $0$  \\
M01V1 & $ 0.63 $ & $33M_\oplus$  & $ 10^{-3}$  \\
M01V2 & $ 0.63 $ & $33M_\oplus$  & $ 3\times 10^{-3}$  \\
\hline
\end{tabular}
\end{center}
\end{table}

\subsection{Circumplanetary disk structure and migration of protoplanets}

Fig.2 shows the circumplanetary disks structure around protoplanets
with mass of $8.5$, $16.5$ and $33M_\oplus$ (top to bottom rows).
The top row panels show the circumplanetary disks surrounding a $8.5
M_\oplus$ protoplanet. In the top row, the left-, middle- and
right-hand panels correspond to models M0026V0 ($\alpha=0$), M0026V1
($\alpha=10^{-3}$) and M0026V2 ($\alpha=3\times 10^{-3}$),
respectively. The middle row panels show the circumplanetary disks
surrounding a $16.5 M_\oplus$ protoplanet. In the middle row, the
left-, middle- and right-hand panels correspond to models M005V0
($\alpha=0$), M005V1 ($\alpha=10^{-3}$) and M005V2 ($\alpha=3\times
10^{-3}$), respectively. The bottom row panels show the
circumplanetary disks surrounding a $33 M_\oplus$ protoplanet. In
the bottom row, the left-, middle- and right-hand panels correspond
to models M01V0 ($\alpha=0$), M01V1 ($\alpha=10^{-3}$) and M01V2
($\alpha=3\times 10^{-3}$), respectively. It can be seen clearly
that as the mass of the protoplanet increases, the spiral structure
of the circumplanetary disk becomes more prominent, higher mass
protoplanet can excite higher amplitude spiral shock.

For the protoplanet with mass of $8.5M_\oplus$, with the increase of
strength of viscosity, the spiral structure is getting weaker. The
spiral structure completely disappears when $\alpha=3\times10^{-3}$
and the circumplanetary disk is nearly axisymmetric about the
protoplanet. For a $8.5 M_\oplus$ protoplanet, the maximum surface
density of the spiral waves is only 2 times bigger than the minimum
surface density; the amplitude of the spiral wave is weak, viscosity
can easily disrupt the spiral structure and make the gas in the
circumplanetary disk smoothly distributed. For the protoplanet with
mass of $16.5M_\oplus$, viscosity also makes the spiral structure of
the circumplanetary disk weaker. But the effect is small compared to
the smaller protoplanet mass case ($M_p=8.5M_\oplus$). This is
because with the increase of the mass of the protoplanet, the
amplitude of the spiral waves gets higher, viscosity can hardly
affect the spiral structure. It can be seen that when the
protoplanet mass reaches $33M_\oplus$, viscosity almost plays no
roles on the circumplanetary disk structure.

The protoplanet excites spiral density waves at the Lindblad
resonance and the torques exerted on the protoplanet as the reaction
of exciting waves make the orbit of the protoplanet around central
star change. Previous works (Lubow et al. 1999; Tanigawa \& Watanabe
2002) found that the spiral structure around the protoplanet can
affect the torques significantly. In the limiting case, if the gas
distribution around a protoplanet is perfectly axisymmetric about
the protoplanet, the torques exerted on the protoplanet will be
zero. From Fig.2, we find that viscosity can disrupt the spiral
structure considerably when the mass of protoplanet $M_p \lesssim
33M_\oplus$. Thus, we can expect that when the protoplanet is small
$(M_p \lesssim 33M_\oplus)$, viscosity can affect the torques
exerted on the protoplanet considerably.

It is always believed that the gas inside the Hill sphere of a
planet migrates with the planet. Therefore, in most previous works,
when calculating the torque exerted on a planet, the contribution
from the gas inside the Hill sphere is completely excluded. However,
Crida et al. (2009) showed that the gas bound to the planet
(circumplanetary disk) only exists inside $0.5 R_H$. In the region
$0.5 R_H < r < R_H$, the gas is not bound to the planet. Therefore,
when calculating the torque, the gas inside the Hill sphere should
not be excluded completely.

Now, we quantitatively study the effects of viscosity on the torques
exerted on a protoplanet. Since we adopt the local approximation,
the net torque in our simulation becomes exactly zero. But an actual
net torque is not zero, owing to slight asymmetry of density
distribution, temperature and the curvature of the protoplanet
orbit. The net torque exerted on the protoplanet is roughly
proportional to the one-side torque exerted by the gas exterior to
the orbit of the protoplanet (Tanigawa \& Watanbe 2002)\footnote{Net
torque is the difference of the two opposite-signed one-side
torques, and the ratio of the two one-side torques does not change
very much when one change parameters such as planet mass or
temperature in linear regime.  Thus, at least in this regime, we can
say that the net torque is proportional to one-side torque.};
therefore, the one-side torque would be a clue to solve the
migration problem. Because we adopt the local approximation, we just
discuss the effects of viscosity on the one-side 'force' exerted on
the protoplanet. The 'force' corresponds to the torque divided by
the semimajor axis of the protoplanet. The one-side force exerted on
a protoplanet is defined by
\begin{equation} F_{y,out}(r_{mask})=\int_0^{x_{max}}
\int_{-y_{max}}^{y_{max}} \Sigma \frac{\partial \Phi}{\partial y}
\theta (r-r_{mask}) dy dx
\end{equation}
where $r_{mask}$ is the artificial inner limit of the force
integration, and $\theta$ is the step function. If $r > r_{mask}$,
$\theta=1$, otherwise $\theta=0$. The 'force density' is defined as
$dF(x)= \int_{-y_{max}}^{y_{max}} \Sigma \frac{\partial
\Phi}{\partial y}  dy $.

\begin{figure}[h!!!]
   \centering
   \includegraphics[width=9.0cm, angle=0]{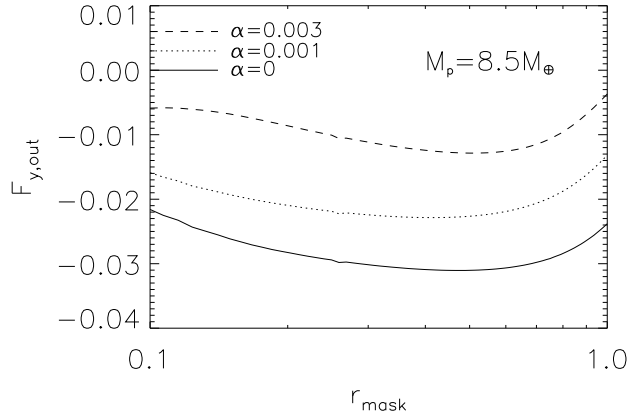}

   \begin{minipage}[]{85mm}

   \caption{Gravitational force exerted on the $8.5M_\oplus$
protoplanet from disk gas in $x>0$ as a function of $r_{mask}$,
within which gas is excluded from force integration. In this figure,
the solid, dotted and dashed lines correspond to models M0026V0
($\alpha=0$), M0026V1 ($\alpha=10^{-3}$) and M0026V2
($\alpha=3\times 10^{-3}$), respectively.}
\end{minipage}
   \label{fig.gammamin}
   \end{figure}

\begin{figure}[h!!!]
   \centering
   \includegraphics[width=9.0cm, angle=0]{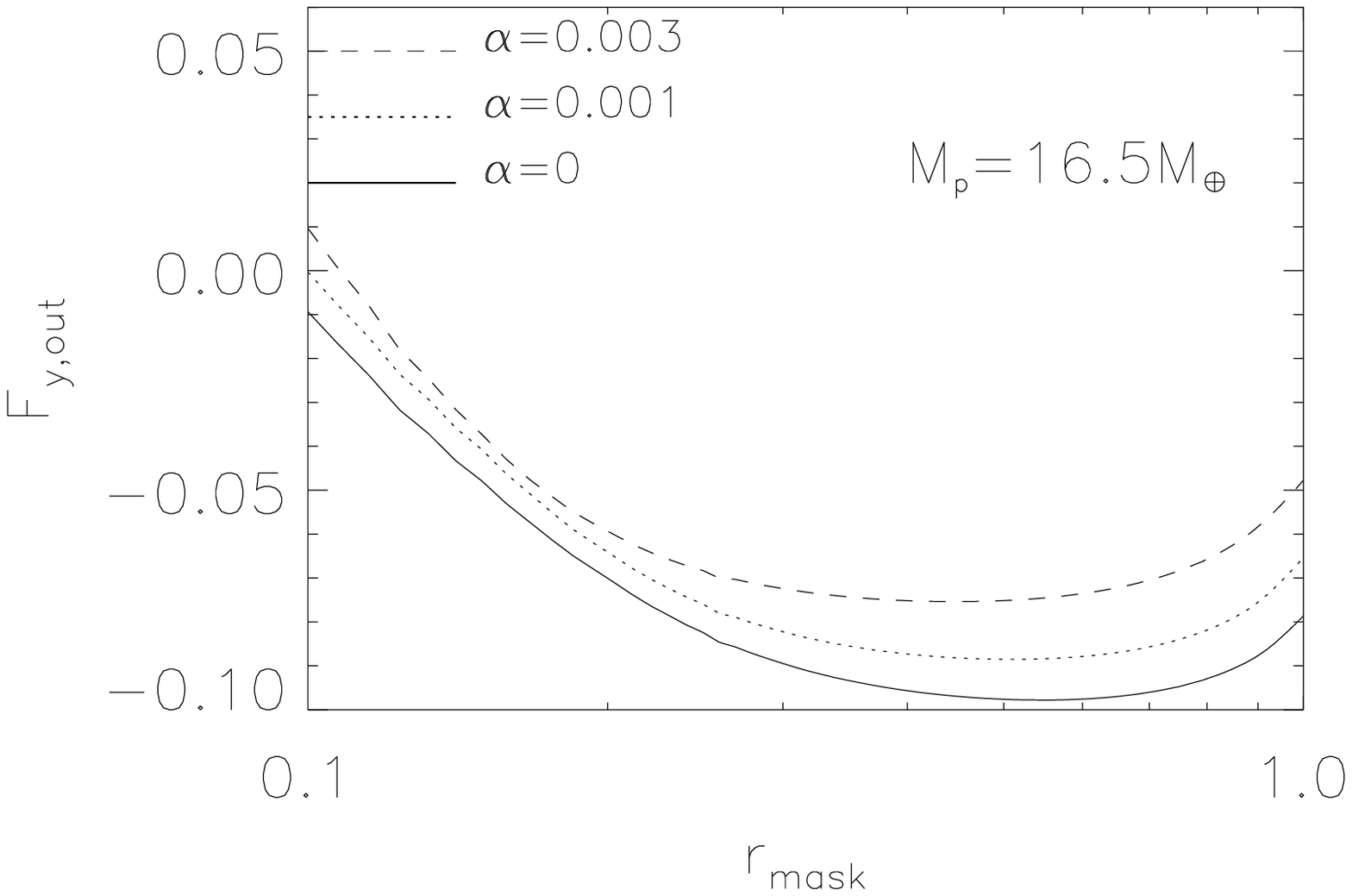}
   \includegraphics[width=9.0cm, angle=0]{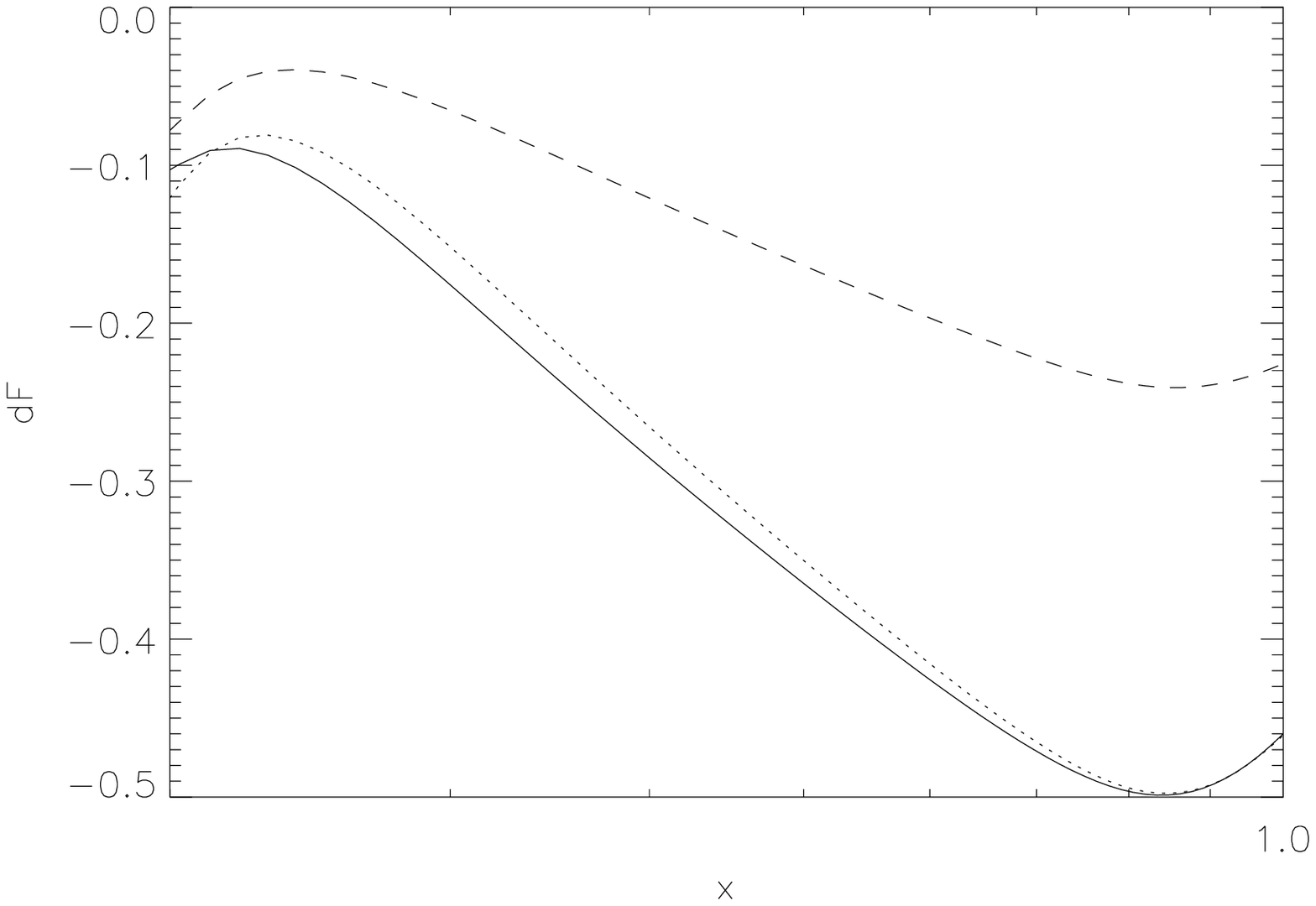}

   \begin{minipage}[]{85mm}

   \caption{Upper-panel shows the gravitational force exerted on the $16.5M_\oplus$
protoplanet from disk gas in $x>0$ as a function of $r_{mask}$,
within which gas is excluded from force integration. In this panel,
the solid, dotted and dashed lines correspond to models M005V0
($\alpha=0$), M005V1 ($\alpha=10^{-3}$) and M005V2 ($\alpha=3\times
10^{-3}$), respectively. Lower-panel shows the torque density as a
function of distance from the $16.5M_\oplus$ protoplanet. In this
panel, the solid, dotted and dashed lines correspond to models
M005V0 ($\alpha=0$), M005V1 ($\alpha=10^{-3}$) and M005V2
($\alpha=3\times 10^{-3}$), respectively.}
\end{minipage}
   \label{fig.gammamin}
   \end{figure}

We show $F_{y,out}$ as a function of $r_{mask}$ for a protoplanet
with mass of $8.5M_\oplus$ in Fig.3. In Section 3.2, we show that
the outer boundary of $8.5M_\oplus$ and $16.5M_\oplus$ protoplanets
located inside $r=0.22$, therefore we can focus on the torque
exerted by the gas beyond $r=0.22$. The gas in $x>0$ exerts a
negative torque on the protoplanet, which pushes the protoplanet
migrating towards the central star. As expected, when the spiral
structure around the planet gets weaker, the force exerted on the
protoplanet is decreased significantly. Fig.4 (upper-panel) plots
$F_{y,out}$ as a function of $r_{mask}$ for a protoplanet with mass
of $16.5M_\oplus$. When $\alpha=3\times 10^{-3}$, the spiral
structure is partially disrupted by viscosity, which also leads to
the reduction of the torque exerted on the planet. In order to show
clearly the 'torque' is affected by viscosity, the lower panel of
Fig.4 plots the 'torque density' as a function of distance from the
planet. The case for a $8.5M_\oplus$ protoplanet is similar. We find
when $M_p\geq 33M_\oplus$, the effect of viscosity on the torque
exerted on protoplanets is negligible. We conclude that the
viscosity can affect the torques exerted on a protoplanets
considerably by disrupting the spiral structure of gas disk when the
mass of the protoplanet is small ($M_p \lesssim 33M_\oplus $).

\begin{figure}[h!!!]
   \centering
   \includegraphics[width=9.0cm, angle=0]{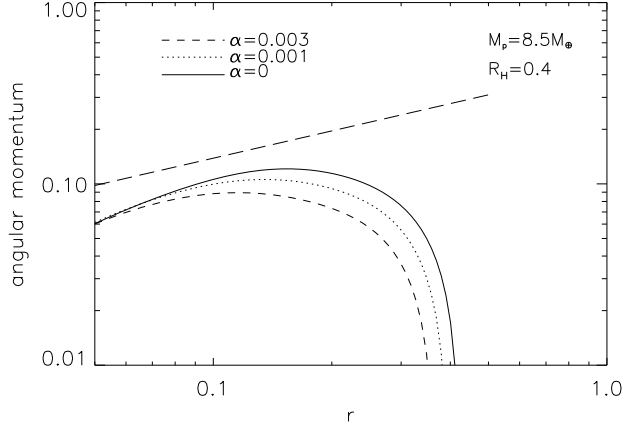}

   \begin{minipage}[]{85mm}

   \caption{Specific angular momentum of the circumplanetary disks
surrounding a $8.5M_\oplus$ protoplanet. In this figure, the solid,
dotted and dashed lines correspond to models M0026V0 ($\alpha=0$),
M0026V1 ($\alpha=10^{-3}$) and M0026V2 ($\alpha=3\times 10^{-3}$),
respectively. The long-dashed line corresponds to the Keplerian
angular momentum with respect to the protoplanet.}
\end{minipage}
   \label{fig.gammamin}
   \end{figure}

\subsection{Circumplanetary disk size}
Now, we investigate the effects of viscosity on the size of the
circumplanetary disks. The radial edge of a circumplanetary disk is
taken as the point of turnover in the specific angular momentum of
the disk. Quillen \& Trilling (1998) made a simple analytic
prediction of the approximate circumplanetary disk radii around
accreting protoplanets. They assume that the gas flows into the Hill
sphere of a protoplanet via the inner and outer Lagrange points.
They also assume that when the gas arrives at the Lagrange points,
the velocity, relative to the Lagrange points, is negligibly small.
The Lagrange points have the same angular velocity around the
central star as the protoplanet. The Lagrange points are
approximately located at
\begin{equation}
r=a_p \pm R_H
\end{equation}

\begin{figure}[h!!!]
   \centering
   \includegraphics[width=9.0cm, angle=0]{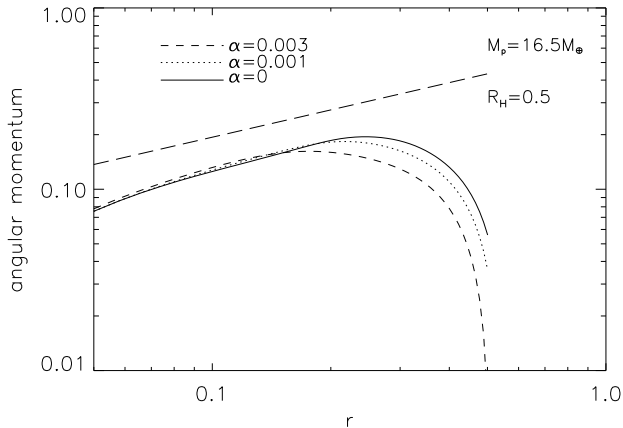}

   \begin{minipage}[]{85mm}

   \caption{Specific angular momentum of the circumplanetary disks
surrounding a $16.5M_\oplus$ protoplanet. In this figure, the solid,
dotted and dashed lines correspond to models M005V0 ($\alpha=0$),
M005V1 ($\alpha=10^{-3}$) and M005V2 ($\alpha=3\times 10^{-3}$),
respectively. The long-dashed line corresponds to the Keplerian
angular momentum with respect to the protoplanet.}
\end{minipage}
   \label{fig.gammamin}
   \end{figure}

Thus, when the gas at the Lagrange points is captured by the
protoplanet, its specific angular momentum relative to the
protoplanet is
\begin{equation}
j\simeq R_H^2 \Omega_p
\end{equation}
Assuming conservation of angular momentum when the accreting gas
flows towards the protoplanet, the centrifugal radius, $r_c$, of the
circumplanetary disk is
\begin{equation}
\frac{j^2}{r_c^2}\approx \frac{GM_p}{r_c}
\end{equation}
which yields \begin{equation} r_c \approx R_H/3. \end{equation}
Thus, the typical size of a circumplanetary disk should be
approximate $R_H/3$.

\begin{figure}[h!!!]
   \centering
   \includegraphics[width=9.0cm, angle=0]{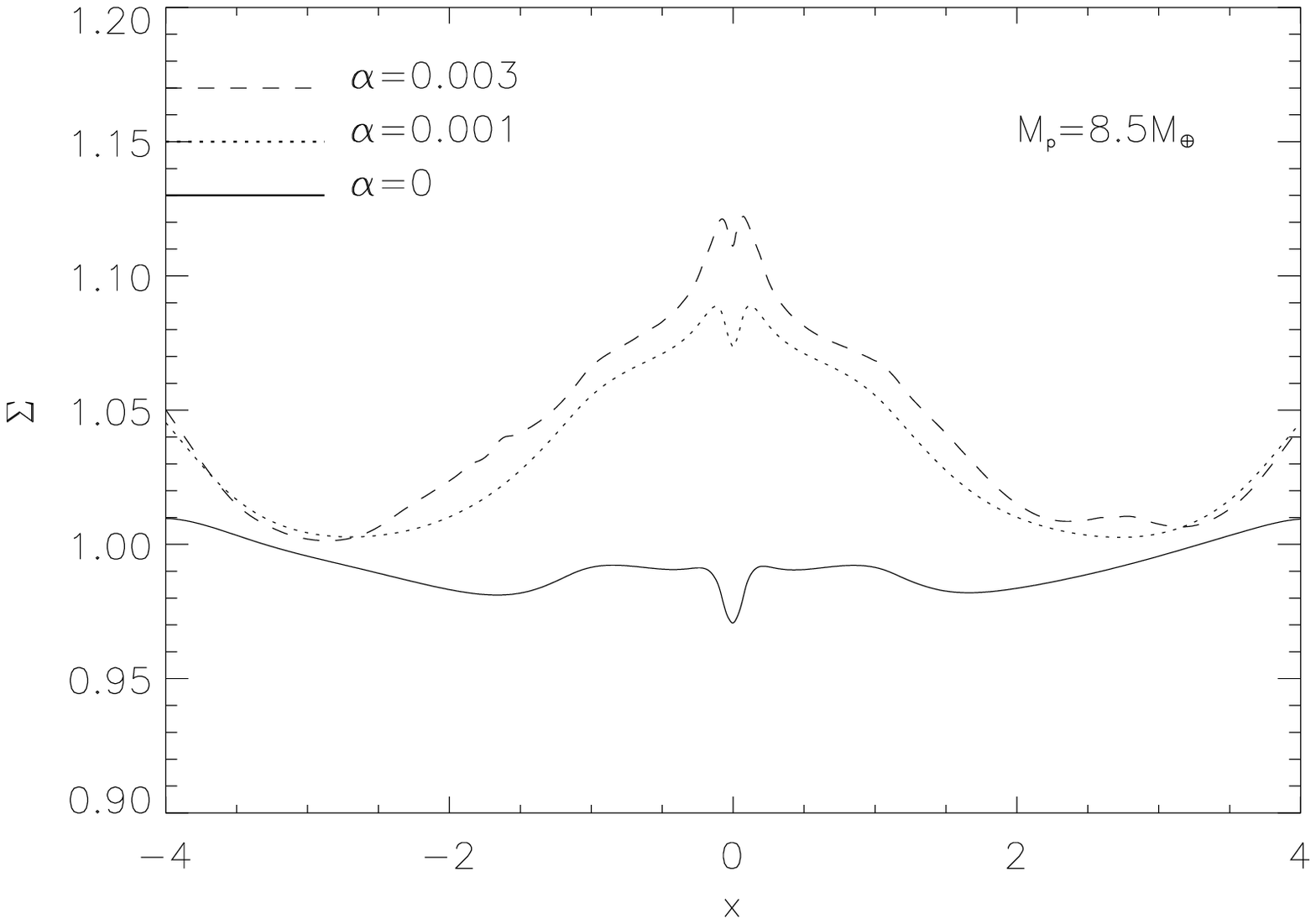}
   \includegraphics[width=9.0cm, angle=0]{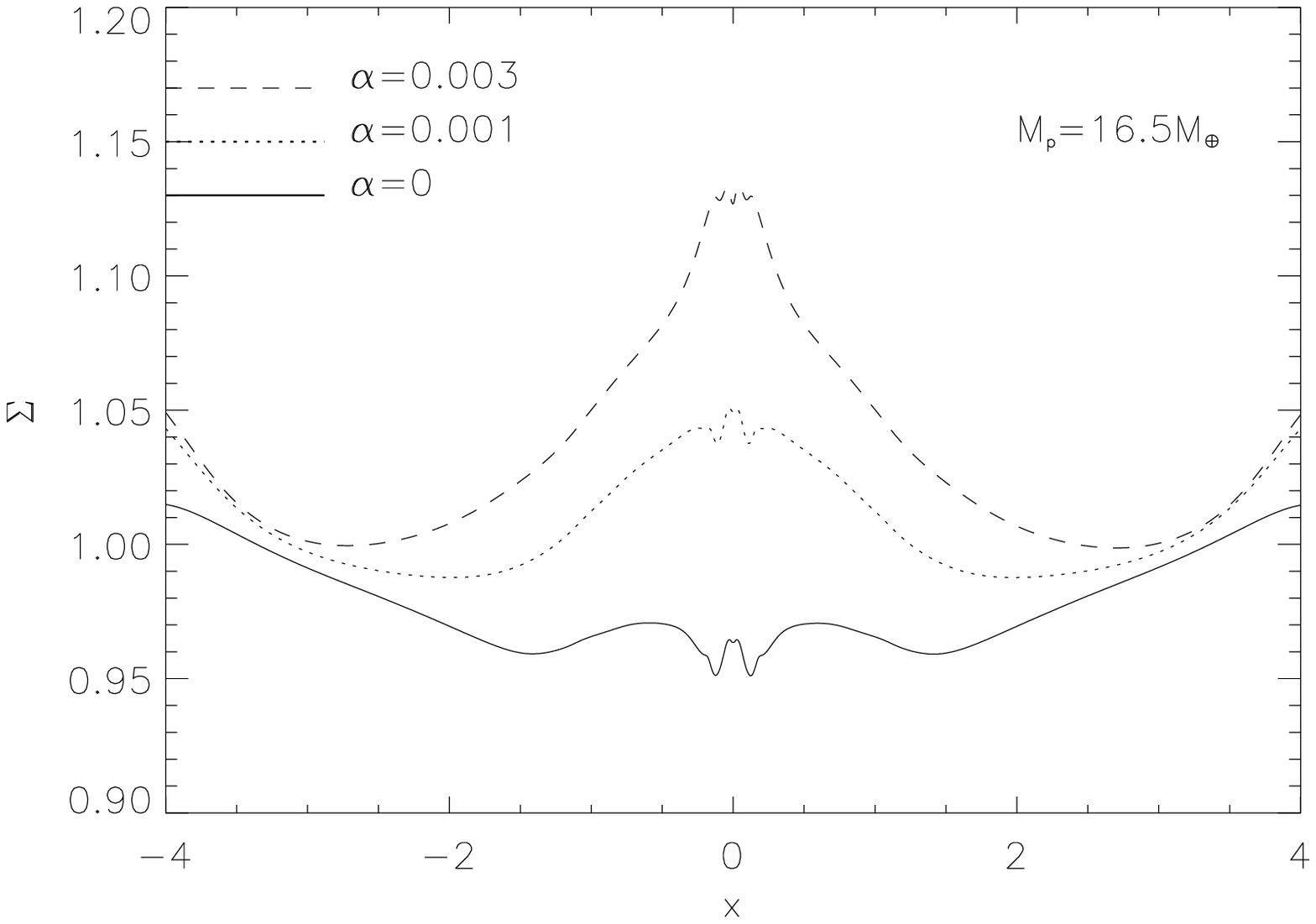}

   \begin{minipage}[]{85mm}

   \caption{The surface density averaged over $y$ against $x$ for the
gas around protoplanets. The upper and lower panels correspond to
$8.5$ and $16.5M_\oplus$ protoplanets, respectively. In the upper
panel, the solid, dotted and dashed lines correspond to models
M0026V0 ($\alpha=0$), M0026V1 ($\alpha=10^{-3}$) and M0026V2
($\alpha=3\times 10^{-3}$), respectively. In the lower panel, the
solid, dotted and dashed lines correspond to models M005V0
($\alpha=0$), M005V1 ($\alpha=10^{-3}$) and M005V2 ($\alpha=3\times
10^{-3}$), respectively.}
\end{minipage}
   \label{fig.gammamin}
   \end{figure}

\begin{figure}[h!!!]
   \centering
   \includegraphics[width=9.0cm, angle=0]{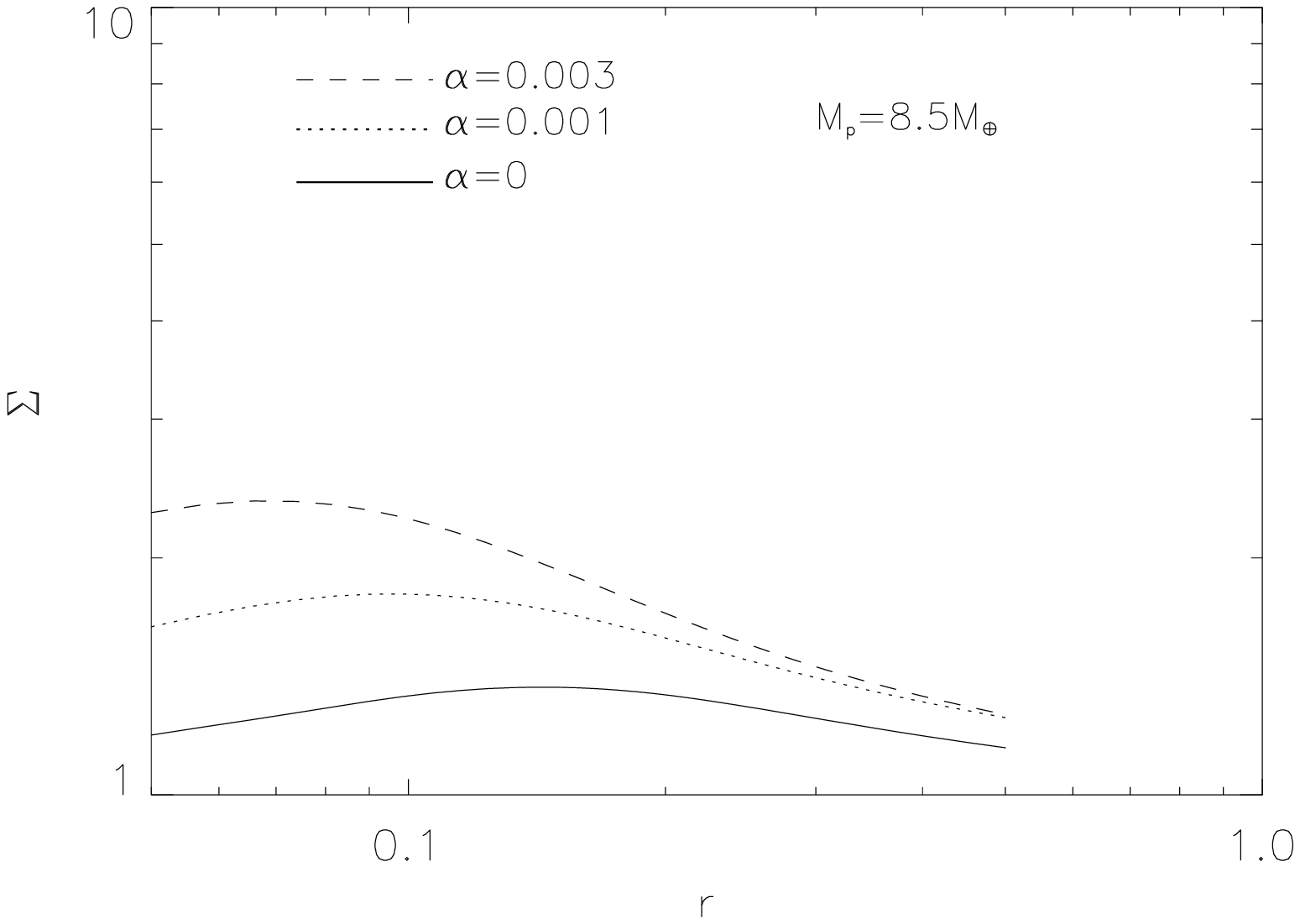}
   \includegraphics[width=9.0cm, angle=0]{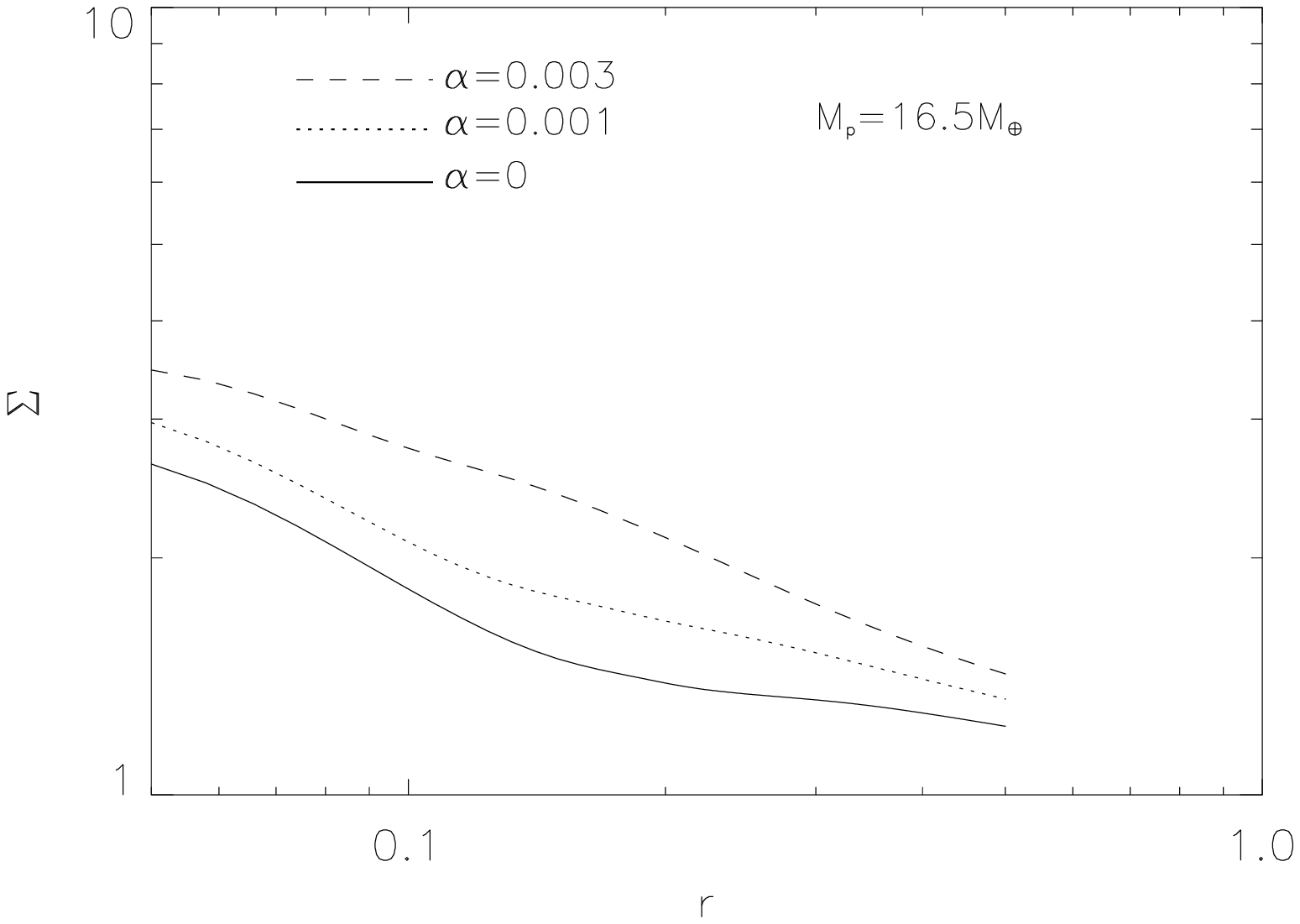}

   \begin{minipage}[]{85mm}

   \caption{Surface density of the circumplanetary disks as a function
of distance $r$ from the protoplanet. The upper and lower panels
correspond to $8.5$ and $16.5M_\oplus$ protoplanets, respectively.
In the upper panel, the solid, dotted and dashed lines correspond to
models M0026V0 ($\alpha=0$), M0026V1 ($\alpha=10^{-3}$) and M0026V2
($\alpha=3\times 10^{-3}$), respectively. In the lower panel, the
solid, dotted and dashed lines correspond to models M005V0
($\alpha=0$), M005V1 ($\alpha=10^{-3}$) and M005V2 ($\alpha=3\times
10^{-3}$), respectively.}
\end{minipage}
   \label{fig.gammamin}
   \end{figure}

Fig.5 shows the specific angular momentum of the circumplanetary
disks around a $8.5M_\oplus$ protoplanet. In this figure, the solid,
dotted and dashed lines correspond to models M0026V0 ($\alpha=0$),
M0026V1 ($\alpha=10^{-3}$) and M0026V2 ($\alpha=3\times 10^{-3}$),
respectively. We can see the outer boundary of a non-viscous
circumplanetary disk (model M0026V0) around a $8.5M_\oplus$
protoplanet locates at $r=0.15$, the disk size does not differ
greatly from the $R_H/3$ prediction. Beyond $r=0.15$, the gas
rotates around the central star, the angular momentum becomes
smaller and smaller with increasing distance from the protoplanet.
The angular momentum relative to the protoplanet even becomes
negative when the distance from the protoplanet is sufficiently big
which is not shown in Fig.5. When viscosity is included, the size of
the circumplanetary disk decreases. In model M0026V2, the outer
boundary of the circumplanetary disk locates at $r=0.11$, the size
of the disk is reduced by a factor of $27\%$ compared to the
non-viscous disk. As can be seen from Fig.5, the specific angular
momentum of the circumplanetary disk becomes lower after including
viscosity, which is due to the outward angular momentum transfer by
viscosity. The variation of angular momentum at given radius is
determined by the divergence of the viscous stress; at the outer
boundary of the circumplanetary disk, the divergence of viscosity is
stronger than that at other radii due to the stronger shear of the
gas (it can be clearly seen from Fig.5), which makes the decreasing
amplitude of the angular momentum at the outer boundary much larger
than that of the gas at other radii. Therefore, the outer boundary
of the circumplanetary disk is moved inward when viscosity is taken
into account.

Fig.6 shows the specific angular momentum of the circumplanetary
disks around a $16.5M_\oplus$ protoplanet. In this figure, the
solid, dotted and dashed lines correspond to models M005V0
($\alpha=0$), M005V1 ($\alpha=10^{-3}$) and M005V2 ($\alpha=3\times
10^{-3}$), respectively. The outer boundary of the circumplanetary
disk is also moved inward when viscosity is included. The outer
boundary of a non-viscous circumplanetary disk (model M005V0) around
a $16.5M_\oplus$ protoplanet locates at $r=0.22$. In model M005V2,
the outer boundary of the circumplanetary disk locates at $r=0.17$,
the size of the disk is reduced by a factor of $23\%$ compared to
the non-viscous disk. We find that when the protoplanet mass $M_p
\gtrsim 33M_\oplus $, the effect of viscosity on the size of a
circumplanetary disk is negligibly small.

Crida et al. (2009) have found that the size of the circumplanetary
disk $\sim 0.5 R_H$, which is slightly larger than that obtained in
this paper. This is because Crida use a energy equation to solve
internal energy of the gas. With the energy equation, the collapse
of the gas onto the circumplanetary disk is limited by the heating
due to adiabatic compression, which gives a wider circumplanetary
disk than in the locally isothermal case.

From Fig. 5 and Fig. 6, we see that the rotational velocity of the
disk around the planets is significantly sub-Keplerian. The reason
is as follows. In the radial direction, gravitational force is
balanced by centrifugal force and pressure gradient force. If the
temperature of the gas is fixed, with the increasing of planet mass,
the centrifugal force will increase (rotational velocity increase).
In our paper, the mass of the planet is small (around 10 earth
mass), so pressure gradient force is important, the gas rotates
sub-keplerian. Tanigawa, Ohtsuki \& Machida (2012) found that the
gas around the planet rotates almost Keplerian. The reason is as
follows. In Tanigawa, Ohtsuki \& Machida, ApJ (2012), the
temperature is identical to that in our paper, but their planet mass
is 1 Jupiter mass which is much bigger than the planet mass in our
paper. Therefore, in Tanigawa, Ohtsuki \& Machida (2012), the gas
rotates almost keplerian.

\subsection{Circumplanetary disk mass}

In the standard picture, a protoplanet residing in a circumstellar
disk can exert torques on the circumstellar disk through the
excitation of spiral density waves (e.g. Goldreich \& Tremaine
1979). The angular momentum carried by the density waves will be
deposited in the circumstellar disk where the waves are damped. The
circumstellar disk gas exterior to the protoplanet orbit gets
positive angular momentum and thus moves outward. The circumstellar
disk gas interior to the protoplanet orbit gets negative angular
momentum and thus moves inward. For a low-mass protoplanet, a
partial density gap forms along the orbit of the protoplanet.
However, the viscosity inside the circumstellar disk tries to
transport gas into the low density partial gap region, which makes
the density gap less prominent. Viscosity can affect the depth of
the density gap around the orbit of the protoplanet and thus affects
the circumplanetary disk mass.

In Fig.7, we plot the surface density averaged over $y$ against $x$
for the gas around protoplanets. The upper and lower panels
correspond to $8.5$ and $16.5M_\oplus$ protoplanets, respectively.
In the upper panel, the solid, dotted and dashed lines correspond to
models M0026V0 ($\alpha=0$), M0026V1 ($\alpha=10^{-3}$) and M0026V2
($\alpha=3\times 10^{-3}$), respectively. In the lower panel, the
solid, dotted and dashed lines correspond to models M005V0
($\alpha=0$), M005V1 ($\alpha=10^{-3}$) and M005V2 ($\alpha=3\times
10^{-3}$), respectively. It is obviously seen that the protoplanets
try to open a density gap along its orbit. As expected, the
protoplanet in the inviscid gas produces a deeper and wider gap than
others.

In Fig.8, we plot the surface density profiles of the
circumplanetary disks around protoplanets. The upper and lower
panels correspond to $8.5$ and $16.5M_\oplus$ protoplanets,
respectively. In the upper panel, the solid, dotted and dashed lines
correspond to models M0026V0 ($\alpha=0$), M0026V1
($\alpha=10^{-3}$) and M0026V2 ($\alpha=3\times 10^{-3}$),
respectively. In the lower panel, the solid, dotted and dashed lines
correspond to models M005V0 ($\alpha=0$), M005V1 ($\alpha=10^{-3}$)
and M005V2 ($\alpha=3\times 10^{-3}$), respectively. Viscosity makes
it easier for gas flows towards the protoplanet. Therefore, the
surface density of the circumplanetary disk in the viscous gas is
higher than that of the inviscid gas. We have calculated the
circumplanetary disk mass. For a $8.5M_\oplus$ protoplanet, when
$\alpha=3\times 10^{-3}$, the disk mass is bigger than the inviscid
disk by a factor of $56\%$. For a $16.5M_\oplus$ protoplanet, when
$\alpha=3\times 10^{-3}$, the disk mass is bigger than the inviscid
disk by an alterative factor of $50\%$. We find that when
$M_p\gtrsim 33M_\oplus$, the effect of viscosity on circumplanetary
disk mass is negligibly small.

\subsection{Mass accretion rate}
We talk about the mass accretion rate in this section. We talk about
mass accretion rate in real unit (equations (11)-(15)). We assume
that the planets locate at $a_p=5.2AU$. Also, we assume that
$M_c=1M_\odot$ and $L=1L_\odot$. In this paper, we find that the
mass accretion rates for non-viscous models M0026V0 ($8.5 M_\oplus$)
and M005V0 ($16.5 M_\oplus$) are $2.2 \times 10^{-5} M_{Jup}/yr$ and
$5.0 \times 10^{-5} M_{Jup}/yr$, respectively. Almost all of the
numerical settings of Tanigawa \& Watanabe (2002) are same as that
in this paper. The only difference is that their is no viscosity in
Tanigawa \& Watanabe (2002). In Tanigawa \& Watanabe (2002), the
smallest mass used is $22 M_\oplus$. Therefore, we can not directly
compare our result to theirs. Fortunately, in equation (19) in
Tanigawa \& Watanabe (2002), they give the dependence of mass
accretion rate on the planet mass. Using their equation (19), the
mass accretion rate for $8.5 M_\oplus$ and $16.5 M_\oplus$ are $2.02
\times 10^{-5} M_{Jup}/yr$ and $4.8 \times 10^{-5} M_{Jup}/yr$,
respectively. Our results for non-viscous gas are consistent with
that in Tanigawa \& Watanabe (2002).

If viscosity is included, we find that the mass accretion rate
becomes higher. The mass accretion rate for the viscous models
M0026V2 ($8.5 M_\oplus$ and $\alpha= 3 \times 10^{-3}$) and M005V2
($16.5 M_\oplus$ and $\alpha= 3 \times 10^{-3}$) are $4 \times
10^{-5} M_{Jup}/yr$ and $8.0 \times 10^{-5} M_{Jup}/yr$,
respectively. D'Angelo et al. (2002) using global simulations to
study the gas flow onto planets. For a $6.4 M_\oplus$ planet, when
$\alpha=4 \times 10^{-3}$, they find that the mass accretion rate is
$1.2 \times 10^{-5} M_{Jup}/yr$ (Fig. 25 in their papers). For a $15
M_\oplus$ planet, when $\alpha=4 \times 10^{-3}$, they find that the
mass accretion rate is $1.5 \times 10^{-5} M_{Jup}/yr$ (Fig. 25 in
their papers). Given that the viscosity and planet mass are
comparable, it seems that the accretion rate found in this paper is
much higher than that obtained in D' Angelo et al. (2002). The
reason may be due to the depletion of the gas inside the orbital
radius of the planet in the global simulations in D' Angelo et al.
(2002).

The Bondi accretion rates for $8.5 M_\oplus$ and $16.5 M_\oplus$
planets located at $5.2 AU$ are $1 \times 10^{-4} M_{Jup}/yr$ and
$3.8 \times 10^{-4} M_{Jup}/yr$, respectively. The angular momentum
of the gas makes the actual accretion rate much smaller than the
Bondi accretion rate of the planets. We must note that the mass
accretion rate obtained in this paper may be not accurate. This is
because, inside the Hill radius, the gas may evolve adiabatically,
the increased temperature towards the planet may make accretion
rates smaller than that obtained in this paper.

\section{Summary and discussion}
We investigate the effects of viscosity on the circumplanetary disks
forming in the vicinity of the protoplanet through two-dimensional
hydrodynamical simulations with the shearing sheet model. We find
that viscosity can affect the properties of the circumplanetary disk
significantly when the mass of the protoplanet $M_p \lesssim
33M_\oplus$.

Local shearing-sheet approximation is only an approximation of the
global model, and it may not be appropriate for investigating the
global evolution of the disk structure. However, Muto et al. (2010)
have shown that the local shearing-sheet approximation and full
global model share many essential physics in common. Muto et al.
(2010) have also shown that the one-dimensional disk evolution model
constructed from the global model and the local model are very
similar. Therefore we expect the local simulations have captured the
main physics of the circumplanetary disks.

Physically we should use three-dimensional simulations to study the
circumplanetary disks inside the Hill radius. However, as a first
step, we carry out these simulations in order to understand the
basic effects of viscosity on the circumplanetary disks. The
two-dimensional simulations can capture the main physics of the
disks. For example, the mass accretion rate can be generated
properly by two-dimensional simulations because the accretion rate
is determined mainly by the flow where $r>R_H$ (when $r>R_H$, the
disk is very thin and two-dimensional simulations is enough)
(Tanigawa \& Watanabe 2002).  Although the 2D simulations can
capture some properties of the circumplanetary disk, some important
feature may lost in the 2D disks. For example, Tanigawa, Ohtsuki \&
Machida (2012) found that most of gas accretion onto circumplanetary
disks occurs nearly vertically toward the disk surface from high
altitude, which can not be found in 2D simulations. In a subsequent
paper, we will discuss the effects of viscosity on the
circumplanetary disks in three-dimensions.

In this paper, we use a smoothing length to smooth the gravity close
to the planet. Muller (et al. 2012) find that for longer distances,
the smooth length is determined solely by the vertical disk
thickness. For the planet case they find that outside $r=H$, when
the value of $r_{sm} = 0.7H$ describes the averaged force very well.
However, for shorter distances the smoothing needs to be reduced
significantly. In this paper, in order to study the structure of
circumplanetary disks we adopt $r_{sm}=0.05H$. This value is proved
to be safe to study the circumplanetary disks by Tanigawa \&
Watanabe (2002).

In this paper, we just use an anomalous stress tensor to mimic the
shear stress, which is in reality magnetic stress associated with
magnetohydrodynamic (MHD) turbulence driven by the magnetorotational
instability (MRI). In a real turbulent circumplanetary disk, the
properties of the disk should fluctuate with time, but we expect
that the time-averaged properties of the turbulent disk should be
consistent with our results here. The spiral shocks in the
circumplanetary disk also can affect the amplitude of the turbulent
stress, but the effect is small when the mass of the protoplanet is
small ($M_p\lesssim 30M_\oplus$) (Papaloizou et al. 2004). Thus, our
calculations have captured the main physics of the circumplanetary
disk.

To properly study the viscous circumplanetary disk, it is necessary
to include magnetic effects. However, the gas in the circumstellar
disk surrounding a protostar is just weakly ionized. Weakly ionized
plasma is subject to a number of non-idea MHD effects due to the
collisional coupling between the ionized species and the neutrals
(e.g. ambipolar diffusion effects) (Bai \& Stone 2011). The
magnetorotational instability (MRI), which is considered as the
major mechanism for angular momentum transport via the MHD
turbulence, is strongly affected by the non-idea MHD effects. Thus,
it is necessary to study the non-idea MHD effects on the MRI before
further investigating the properties of the magnetized
circumplanetary disks.

We find that when $M_p \lesssim 33M_\oplus$, viscosity can disrupt
the spiral structure around a protoplanet considerably and make the
gas in the disk smoothly distributed, which makes the torques
exerted on the protoplanet weaker. Thus, viscosity can make the
migration speed of a protoplanet lower. This is helpful to solve the
problem that a protoplanet quickly migrates to the vicinity of the
central star before becoming a gas giant planet.

According to the core accretion theory, the formation process of a
gas giant planet can be divided into three phases. (1) In phase 1,
the solid core forms first, which has a mass of several Earth mass.
(2) In phase 2, a spherically hydrostatic gas envelope around the
solid core forms. The gas accretion rate in this phase is very low.
(3) After the point when the core mass and the envelope mass become
comparable, gas is accreted in a runaway fashion. The main problem
of the core accretion theory is that the formation time of a gas
giant planet exceeds the lifetime of the circumstellar disk. Under
the assumption of spherical symmetry and gas in hydrostatic
equilibrium, previous works found that the time needed to complete
phase 2 is comparable or exceeding the lifetime of the circumstellar
disk (e.g. Pollack et al. 1996). The reason for the longtime
evolution in phase 2 is that they assume spherical accretion. The
gravitational energy released in the accretion process can not
easily escape from the system, the thermal pressure can support the
envelope against the gravity of the core, which decreases the
accretion rate in phase 2 significantly. Lin (Lin 2006) proposed
that if a circumplanetary disk (instead of a spherically hydrostatic
envelope) exists around a several Earth mass protoplanet, the
evolution time of phase 2 may be decreased significantly. This is
because in the disk accretion case, the gravitational energy
released in the accretion process can easily escape from the surface
of the circumplanetary disk. However, our simulation found that when
the protoplanet is small (several Earth mass), the size of the
circumplanetary disk is just $\sim 27\%$ of the Hill radius, $R_H$,
when viscosity is included (see Fig.4). Thus, even though the
existence of a circumplanetary disk can make the formation time of a
gas giant planet smaller, we expect that the effect is not important
because the size of the disk is small enough compared to the Hill
radius.

Satellites may form in the circumplanetary disk. Regular satellites
around a gas giant planet are considered to form according to a
scenario similar to the formation of Earth-like planets in our Solar
system. The protosatellite forms by the accumulation through mutual
collision of the satellitesimals that form after the dust grains
sink towards the equatorial plane of the circumplanetary disk (e.g.
Stevenson, Harris \& Lunine 1986). After including viscosity, the
density of the circumplanetary disk increases, which should be
helpful for the formation of satellitesimals by the accumulation of
dust grains. However, the increase in density will result in the
increase of opacity. It is difficult for high opacity
circumplanetary disks to radiate its energy generated by gas
accretion. Thus, the increase of density may result in the increase
in temperature of the circumplanetary disks. Large fraction of ice
is found in some of the Galilean moons around the Jupiter (Schubert,
Spohn \& Reynolds 1986). This means that the temperature of the
circumplanetary disk should be low enough for water condensation
into ice (Canup \& Ward 2002). In this sense, the increase of
density may make it difficult for satellite formation. In order to
understand the properties of the circumplanetary disk better, we
need to study the viscous circumplanetary disk with radiative
transfer, which is beyond the scope of this paper.

\normalem
\begin{acknowledgements}
We thank Ruobing Dong and T. Tanigawa for useful discussions. We
thank Hsiang-Hsu Wang for numerical supports. This work was supported in part by the Natural Science Foundation of China (grants 10833002, 10825314, 11103059, 11121062, and 11133005), the National Basic Research Program of China (973 Program 2009CB824800), and the CAS/SAFEA International Partnership Program for Creative Research Teams. The simulations were carried out at Shanghai Supercomputer Center.
\end{acknowledgements}

\label{lastpage}

\end{document}